\documentclass[
    ,draft            
  ]
  {aipproc}

\layoutstyle{6x9}

\usepackage{bm}

\begin{document}

\title{Generalized parton distributions:
Status and perspectives}


\author{C.~Weiss}{
  address={Theory Center, Jefferson Lab, Newport News, VA 23606, USA}
}

\begin{abstract}
We summarize recent developments in understanding the concept of 
generalized parton distributions (GPDs), its relation to nucleon 
structure, and its application to high--$Q^2$ electroproduction processes. 
Following a brief review of QCD factorization and transverse nucleon 
structure, we discuss (a) new theoretical methods for the analysis of 
deeply--virtual Compton scattering ($t$--channel--based GPD 
parametrizations, dispersion relations); (b) the phenomenology 
of hard exclusive meson production (experimental tests of 
dominance of small--size configurations, model--independent 
comparative studies); (c) the role of
GPDs in small--$x$ physics and $pp$ scattering (QCD dipole model,
central exclusive diffraction). We emphasize the usefulness
of the transverse spatial (or impact parameter) representation 
for both understanding the reaction mechanism in hard exclusive processes 
and visualizing the physical content of the GPDs.
\end{abstract}

\maketitle

\section{Introduction}
Generalized parton distributions (GPDs) have established themselves
as a key concept in the study of hadron structure in QCD.
On the experimental side, they enable a unified description of
several high--$Q^2$ electroproduction processes (inclusive, exclusive)
on the basis of QCD factorization. On the theoretical side, 
they provide us with information about the transverse spatial distribution
of quarks/gluons and have given rise to new ideas of
``quark/gluon imaging'' of the nucleon. Studies of GPDs are
the subject of completed and on-going experiments (HERA, HERMES, JLab 6 GeV) 
and are an essential part of the physics program of future facilities for
$ep/eA$ scattering (JLab 12 GeV, EIC). The purpose of this article is
to summarize the status and future perspectives of this field,
with emphasis on the interplay between experimental data and 
theoretical/phenomenological concepts; for more detailed reviews, see 
Ref.~\cite{Goeke:2001tz}. 
\section{QCD factorization in high--$Q^2$ electroproduction}
\begin{figure}
\includegraphics[width=.95\textwidth]{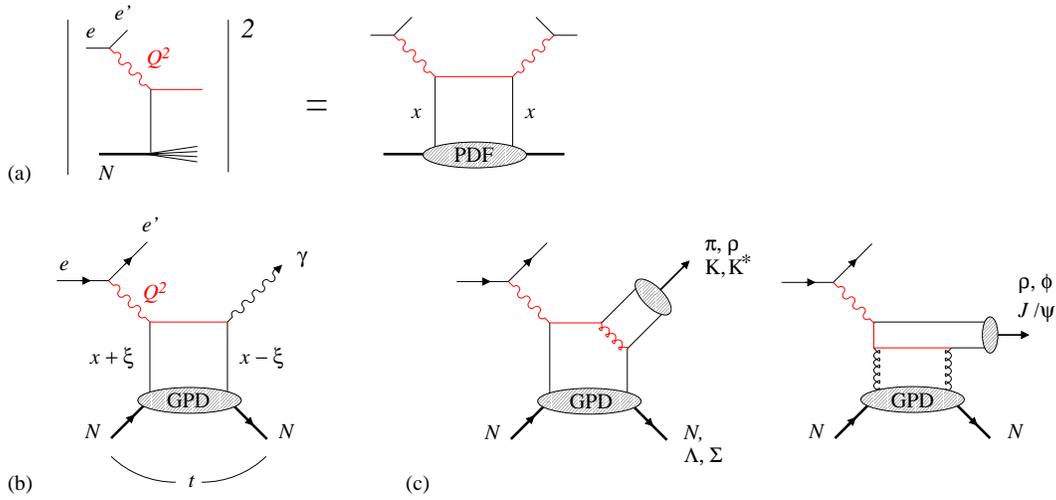}
\caption{Factorization in high--$Q^2$ electroproduction.
(a) Inclusive production $N(e, e')X$. (b, c) Exclusive production 
$N(e, e'M)N \; (M = \gamma, \textrm{meson})$. The parameter $\xi$ is 
related to the Bjorken scaling variable, $\xi = x_B / (2 - x_B)$,
and kinematically fixed (longitudinal momentum transfer to target).}
\label{fig:fact}
\end{figure}

The basis for a partonic description of hadron structure is the 
idea of factorization in high--$Q^2$ electroproduction processes.
At $Q^2 \gg R_{\rm had}^{-2}$ the virtual photon has a transverse
resolution much smaller than the typical hadronic size, and the 
scattering process takes place with a quasi--free parton
in the nucleon. This is well--known in inclusive DIS 
(see Fig.~\ref{fig:fact}a), where the cross
section can be factorized into that of the parton--level
process, involving transverse distances $\sim 1/Q$, and the distributions 
of partons in the nucleon (PDFs), governed by long--distance dynamics 
($\sim R_{\rm had}$). The latter are formally defined as the 
expectation values of a certain quark bilinear operators in the nucleon state, 
$\langle N| \bar\psi(0) \ldots \psi (z) | N \rangle_{z^2 = 0}$, and can 
be interpreted in space--time as the densities of quarks with longitudinal
momentum $xP$ in a fast--moving nucleon (momentum $P \gg R_{\rm had}^{-1}$).
%
%

The same reasoning applies to certain exclusive production processes
at high $Q^2$, specifically deeply--virtual Compton scattering (DVCS) and 
exclusive meson production (see Fig.~\ref{fig:fact}b and c), where
the final--state photon (meson) is produced in the reaction with a 
single parton in the nucleon. The amplitudes can be factorized into
the quark--level production amplitude and the GPDs, describing the
coupling of the active parton to the nucleon.
They depend on the momentum fractions of the initial and final parton,
$x\pm \xi$ ($-2\xi$ is the fractional longitudinal momentum transfer 
to the nucleon), as well as the transverse momentum 
transfer to the nucleon, $\bm{\Delta}_\perp$, or, alternatively, the 
invariant momentum transfer, $t$. The GPDs are formally defined
as the transition matrix elements $\langle N'| \ldots |N\rangle$
of the quark (or gluon) bilinear operators and combine aspects of the parton 
densities with those of the nucleon elastic form factors; in fact, 
they contain both as limiting cases. QCD factorization guarantees that they
are universal, process--independent characteristics of the nucleon,
which can be probed using different partonic scattering processes 
(\textit{e.g.}\ different meson production channels); this fact is 
essential not only for experimental studies of GPDs, but also for 
our ability to calculate them using non-perturbative QCD methods
such as lattice simulations. (For a discussion of QCD radiative 
corrections and the scale dependence of GPDs, see Ref.~\cite{Goeke:2001tz}.)

QCD factorization is formally justified is the $Q^2 \rightarrow \infty$
limit and assumes that the production process takes place predominantly 
over transverse distances $\sim 1/Q$. In processes at $Q^2 \sim 
\textrm{few GeV}^2$ there can be significant contributions from 
finite--size configurations, leading to power ($1/Q^2$, higher twist)
corrections to the amplitude, in particular in meson production 
(see below). Combining small--size and finite--size contributions
in a consistent approach is an important problem for phenomenological studies.

\section{$\textrm{GPDs}$ and nucleon structure}
\begin{figure}[b]
\includegraphics[width=.99\textwidth]{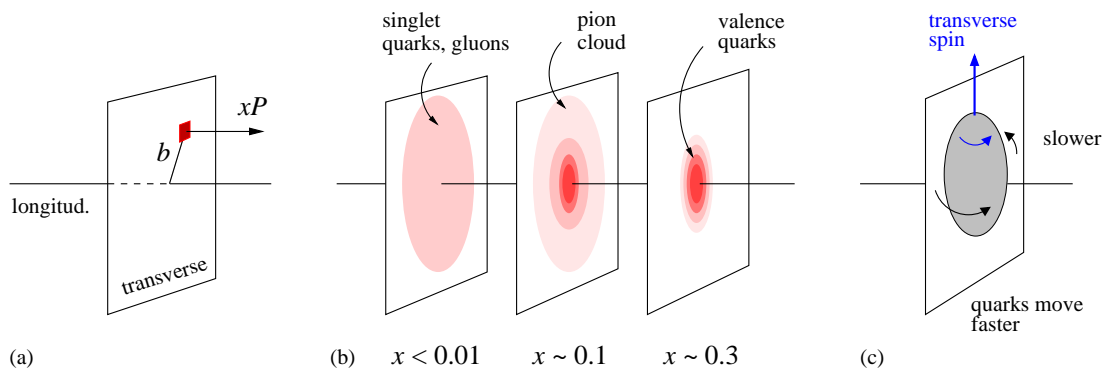}
\caption{(a) Impact parameter representation of the GPDs ($\xi = 0$).
(b) Tomographic parton images of the nucleon at fixed longitudinal 
momentum fraction $x$. (c) Distortion of the longitudinal motion 
of partons by transverse polarization of the nucleon.}
\label{fig:tomogr}
\end{figure}

The concept of GPDs has proved to be a most useful tool in the general
effort to describe nucleon structure in terms of QCD degrees of freedom. 
The most interesting aspect is that the GPDs describe the transverse 
spatial distribution of quarks and gluons in the nucleon. 
This is seen most easily in the special case of zero longitudinal
momentum transfer, $\xi = 0$, where the GPD represents the ``transverse 
form factor'' of the QCD operator measuring the density of partons with 
longitudinal momentum fraction $x$. Its Fourier transform 
$\bm{\Delta}_\perp \rightarrow \bm{b}$ describes the distribution 
of these partons with respect to their transverse displacement, 
$\bm{b}$, from the center--of--momentum of the nucleon
(impact parameter representation, see Fig.~\ref{fig:tomogr}a) 
\cite{Burkardt:2000za}. It corresponds to a set of tomographic parton
images of the nucleon at fixed longitudinal momentum fraction $x$ 
(Fig.~\ref{fig:tomogr}b). This representation provides a natural framework 
for visualizing the partonic structure and discussing
its relation to long--distance hadronic dynamics. 
At $x \sim 0.3$ on ``sees'' the valence
quarks (and gluons), which are distributed over transverse distances 
$b \ll 1 \, \textrm{fm}$. At $x < M_\pi/ M_N = 0.15$ the
pion cloud makes a distinct contribution to the partonic structure,
extending up to transverse distances $b \sim M_\pi^{-1}$
\cite{Strikman:2003gz}. At even smaller $x$ the dominant partons
are radiatively generated singlet quarks and gluons; the
area of their transverse distribution $\langle b^2 \rangle$ grows
logarithmically with $1/x$ (effective Regge slope).
%
%

The transverse coordinate representation also lends itself to
the discussion of polarization effects \cite{Burkardt:2002hr,Diehl:2005jf}. 
There are 4 helicity components of the quark GPDs, corresponding to the
operators measuring the sum/difference of quark helicities 
(\textit{cf.}\ unpolarized/polarized PDFs) and their nucleon 
helicity non-flip/flip matrix elements (\textit{cf.}\ Dirac/Pauli
and axial/pseudoscalar form factors). The Pauli form factor--type 
GPD ($E$) has an interesting interpretation in the transverse nucleon 
spin basis; it describes the distortion of the longitudinal motion
of the quarks in a transversely polarized nucleon 
(Fig.~\ref{fig:tomogr}c) \cite{Burkardt:2002hr}. Likewise, for a 
longitudinally polarized nucleon, one can form combinations of the
unpolarized and polarized GPDs corresponding to definite quark 
helicity states ($H \pm \tilde H$), and study the spatial 
distribution of quarks polarized along or opposite to the nucleon spin.

Another reason for interest in the GPDs is that their moments
are related to fundamental static properties of the nucleon. 
Integration over $x$ with a weighting factor $x^{n-1}$ projects out 
the spin--$n$ component of the non--local quark bilinear operator. 
The spin--2 component coincides with the quark part of the QCD 
energy--momentum (EM) tensor, whose matrix elements cannot be measured 
directly with local electroweak probes. In particular, the EM tensor 
is related to the angular momentum operator; this is the basis of 
the Ji sum rule which expresses the total quark angular momentum
in terms of the second moments of the unpolarized GPDs $E$ and $H$
\cite{Ji:1996ek,Voutier:2009sc}. The components of the EM tensor
also provide information about the QCD forces acting on a quark
in the nucleon at rest (pressure, shear forces) \cite{Polyakov:2002yz}; 
these relations have recently been illustrated by a calculation 
within the chiral quark--soliton model of the nucleon \cite{Goeke:2007fp}.

Moments of the GPDs, as matrix elements of local operators, can also be 
calculated in lattice QCD \cite{Hagler:2007xi,Zanotti:2008zm}. 
Present simulations are limited to
$n \leq 4$, allowing for first conclusions
about the correlation of $x$-- and $t$--dependence in non-singlet 
GPDs (singlets require inclusion of disconnected diagrams). Accurate 
lattice results would have the potential to constrain future
GPD parametrizations.

\section{$\textrm{GPDs}$ in $eN$ scattering}
Quantitative studies of GPDs draw from several sources of information.
Important basic information comes from the parametrizations of PDFs 
and elastic nucleon form factors, which, respectively, constrain the 
GPDs in the zero momentum--transfer limit ($\xi = 0, t = 0$) and their 
first moments as functions of $t$. The ``new'' information about the
correlation of the $x, \xi$ and $t$--dependence, which \textit{e.g.}\
determines the transverse profile of the nucleon and its
change with $x$ in Fig.~\ref{fig:tomogr}b, comes from measurements
of high--$Q^2$ exclusive processes, in particular DVCS and
meson production (see Fig.~\ref{fig:fact}b and c).

The analysis of high--$Q^2$ exclusive processes 
generally proceeds in two steps. One first investigates 
the reaction mechanism and tries to ascertain that a description
based on short--distance dominance and QCD factorization is applicable; 
this is done by testing certain qualitative implications of the
approach to the short--distance regime ($Q^2$--scaling, \textit{etc}.) 
which do not depend on the specific form of the GPDs. 
One can then attempt to extract quantitative information 
about the GPDs by analyzing the kinematic dependences of the 
leading--twist (leading in $1/Q^2$) observables. 
The details of this program vary considerably 
between DVCS and meson production, because of the different complexity 
and reaction dynamics of the two processes.

\begin{figure}
\includegraphics[width=.98\textwidth]{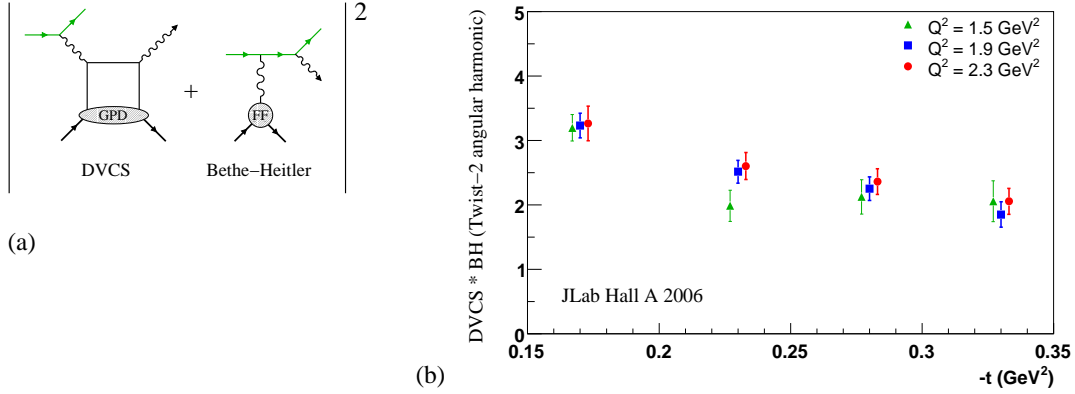} 
\caption{(a) Interference of DVCS and BH processes
in $N(e, e'\gamma )N$. (b) The beam spin--dependent BH-DVCS interference 
cross section (Twist--2 angular harmonic) measured in the JLab Hall A 
experiment \cite{Munoz Camacho:2006hx}, 
as a function of $t$, for several values of $Q^2$.}
\label{fig:dvcs}
\end{figure}

\textit{Deeply--virtual Compton scattering.}
DVCS is generally regarded as the most promising channel for
probing GPDs in the valence quark regime. In this
process one expects short--distance dominance to set in already
at $Q^2 \sim \textrm{few GeV}^2$, based on the formal analogy
of DVCS to inclusive DIS and the experience with other 
exclusive two--photon processes, such as $\gamma^\ast\gamma 
\rightarrow \pi^0$ studied in $e^+e^-$ annihilation.
In the measured $N(e, e'\gamma)N$ cross section the DVCS process
interferes with the Bethe--Heitler (BH) process, whose amplitude is
purely real and calculable in terms of the nucleon elastic 
form factors (see Fig.~\ref{fig:dvcs}a);
this circumstance amplifies the DVCS amplitude and allows one to
measure it --- and thus the GPDs --- at the amplitude level, by isolating
the interference term through polarization observables.

Recent measurements of the beam spin--dependent BH--DVCS 
interference cross section in the JLab Hall A 
experiment \cite{Munoz Camacho:2006hx}
indicate an early approach to $Q^2$--scaling (see Fig.~\ref{fig:dvcs}b),
in accordance with theoretical expectations, suggesting
that a description based on leading--twist QCD factorization
should be applicable already at $Q^2 \sim \textrm{few GeV}^2$. 
Present experimental efforts focus on
separating the BH--DVCS interference and the $\textrm{DVCS}^2$ 
term in the beam spin--independent cross section, separating
the nucleon helicity--components of the GPDs through measurement
of target spin asymmetries, and measuring neutron GPDs through
quasi--free neutron DVCS with nuclear targets \cite{Voutier:2009sc}.
%
%

The methods for analyzing DVCS observables at the leading--twist 
level and extracting information about GPDs are well developed 
and have been extensively discussed in the literature. Recently, 
two important new tools were added to the arsenal.
One is GPD parametrizations based on the idea of a $t$--channel 
partial--wave expansion of the hard exclusive amplitude. In the 
so--called dual parametrization \cite{Polyakov:2002wz}, inspired 
by the analogy with dual amplitudes in hadron--hadron scattering, 
the GPD \textit{viz.}\ hard exclusive amplitude is represented 
as a formal series of $t$--channel resonance exchanges, which is 
then resummed and parametrized in terms of functions of the complexity 
of usual parton densities (``forward--like'' functions; 
see Ref.~\cite{SemenovTianShansky:2008mp} for a recent summary). 
In the approach of Ref.~\cite{Mueller:2005ed}, based on the analogy 
with Regge theory, a similar representation is derived using complex 
angular momentum techniques. These $t$--channel representations have 
several advantages over the traditional double distribution
(spectral) GPD parametrization \cite{Radyushkin:1997ki}: 
(a) they diagonalize the QCD evolution equations for GPDs by 
using the conformal moments of the QCD operator; (b) they suggest a 
natural high--energy (small--$\xi$) expansion of the hard 
exclusive amplitudes. The representation of Ref.~\cite{Mueller:2005ed}
has successfully been applied to fit the HERA DVCS data at NLO 
accuracy \cite{Kumericki:2007sa}; for a summary of the status of 
LO fits based on the dual parametrization see Ref.~\cite{Guzey:2008ys}.
A simple GPD model based on the dual parameterization describes 
the bulk of the JLab spin--dependent cross section and asymmetry
data \cite{Polyakov:2008xm}. However, as a general GPD parametrization
the $t$--channel representation is likely to be less useful in the
large--$x_B$ (large--$\xi$) region.

The other new tool are $s$--channel dispersion relations for the DVCS
amplitude \cite{Teryaev:2005uj}; see also Ref.~\cite{Kumericki:2007sa}.
They allow one to restore the real part of the DVCS amplitude from the 
imaginary part, which is directly accessible in the spin--dependent 
BH--DVCS interference cross section, and an energy--independent
subtraction constant. In the GPD description, the imaginary part
is proportional to the GPD at the transition points $x = \pm \xi$,
while the real part is given by a certain integral of the GPD 
over $x$. The existence of a dispersion relation thus implies that 
there is a ``hidden'' relation between the values of the GPDs
at $x = \pm \xi$ and elsewhere; this property is in a subtle way
related to the fact that moments of the GPD, as matrix elements of
local spin--$n$ operators, are required by Lorentz invariance
to be polynomials of degree $n$ in $\xi$ (``polynomiality condition'').
The subtraction constant turns out to be related to the
D--term \cite{Polyakov:1999gs}, a distinct component of the
unpolarized GPDs required by polynomiality, which is located in the region
$-\xi < x < \xi$ and drops out in the zero momentum--transfer limit 
($\xi = 0$), and thus has no analog in the usual parton densities. 
If confirmed by further studies, these recent findings 
would have profound implications for 
the GPD analysis of DVCS data in the valence quark regime ($x_B > 0.1$). 
They would mean that in measurements at fixed $Q^2$ only the GPDs 
at $x = \pm\xi$ (as functions of $\xi$ and $t$) and the D--term 
(as a function of $t$ alone) are observable; whatever information
from the GPDs at $x \neq \pm\xi$ enters in the amplitude could be 
reconstructed from the dispersion relation. 
The main task of experiments then would be to provide 
accurate and comprehensive data on the imaginary part of the
DVCS amplitude through measurements of the spin--dependent 
interference cross section. 

\textit{Exclusive meson production.} In meson production the
quantum numbers of the produced meson determine the spin 
(unpolarized/polarized), charge parity ($q\pm \bar q$), and
flavor of the GPD, allowing one to probe the individual GPD components
more selectively than in DVCS. However, because of the additional 
interaction required in the formation of the meson the partonic 
process is more complicated, and detailed study of the reaction 
mechanism is necessary before a factorized description 
can be applied.

\begin{figure}
\includegraphics[width=.9\textwidth]{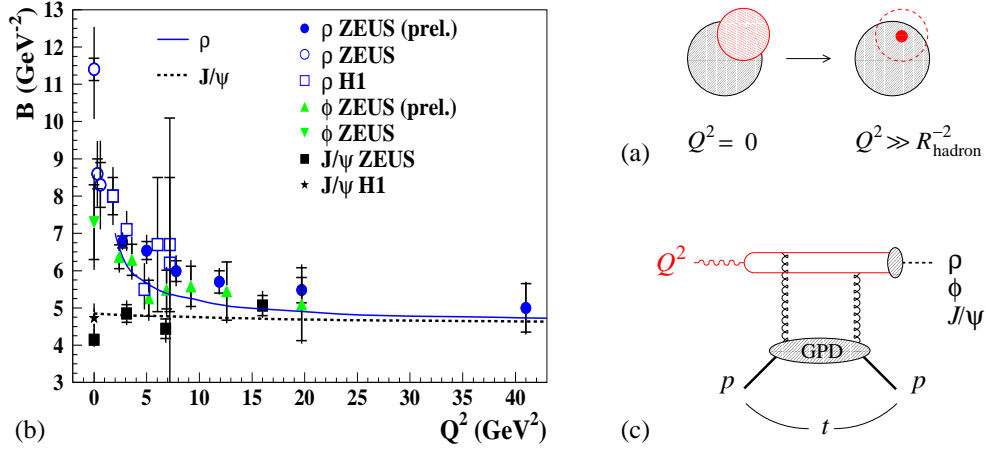} 
\caption{(a) Approach to the small--size regime in meson production
(transverse view). (b) The exponential $t$--slope, $B$, for several 
vector meson electroproduction channels measured at HERA, as a function 
of $Q^2$ \cite{Frankfurt:2005mc}. (c) In the small--size regime, 
the measured $t$--dependence can be identified with that 
of the gluon GPD.}
\label{fig:tslope}
\end{figure}

QCD factorization for meson production relies on the fact that
for $Q^2 \rightarrow \infty$ the meson is predominantly produced 
in a small--size configuration (transverse size $\sim 1/Q$), whose
coupling to the target is weak (``color transparency'') and 
can be computed perturbatively. The approach to the small--size
regime with increasing $Q^2$ can be verified experimentally
in a model--independent manner and constitutes a crucial test of
the reaction mechanism. The $t$--slope of the differential cross
section (more precisely, the $\bm{\Delta}_\perp^2$ slope) measures
the transverse area of the interaction region, reflecting the
size of the target and that of the dominant configurations
of the produced meson. As $Q^2$ increases and small--size meson
configurations become more important, one expects the $t$--slope to 
decrease and eventually stabilize (see Fig.~\ref{fig:tslope}a).
Exactly this behavior is seen in the HERA vector meson
production data (see Fig.~\ref{fig:tslope}b), 
where the exponential $t$--slope of $\rho^0$ and 
$\phi$ production changes from the ``soft'' value of 
$B \sim 10 \, \textrm{GeV}^2$ at $Q^2 = 0$ to a value of 
$B \sim 5 \, \textrm{GeV}^2$ at $Q^2 \sim 20-30 \, \textrm{GeV}^2$;
the latter value is close to the slope of $J/\psi$ production,
which is practically $Q^2$--independent because the $J/\psi$
is produced in small--size configurations even at $Q^2 = 0$.
These observations attest to the approach to the
small--size regime in vector meson production at high $Q^2$.
The limiting value of the $t$--slope for $Q^2 \rightarrow \infty$ 
then reflects the size of the target only and can be associated 
with the $t$--dependence of the GPD (see Fig.~\ref{fig:tslope}c). 

%
%

\begin{figure}
\includegraphics[width=.9\textwidth]{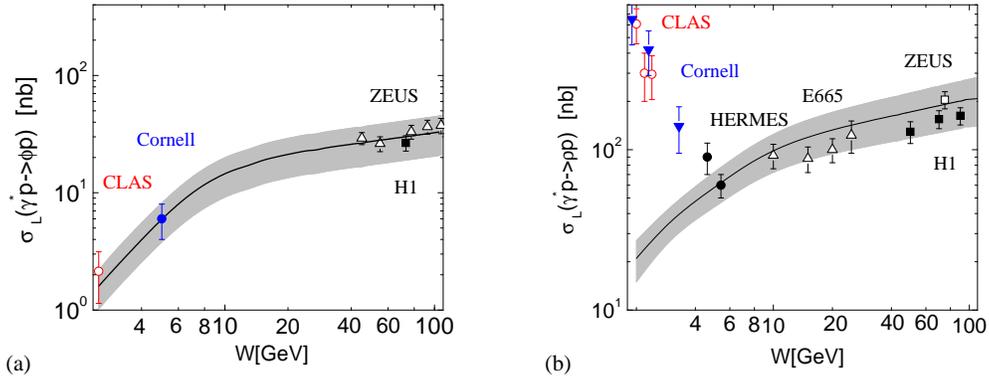} 
\caption{The longitudinal cross section for (a) exclusive $\phi$,
(b) exclusive $\rho^0$ production (adapted from Ref.~\cite{Goloskokov:2005sd}).
The curve/error band show the GPD--based model 
calculation of Ref.~\cite{Goloskokov:2005sd}.}
\label{fig:kroll}
\end{figure}

In vector meson production at $Q^2 \sim \textrm{few GeV}^2$ the
data indicate substantial contributions from  finite--size configurations
in the produced meson. This is confirmed by theoretical 
calculations which account for the finite meson size by including the 
intrinsic quark transverse momentum in the partonic scattering process 
(higher--twist corrections) 
\cite{Frankfurt:1995jw,Vanderhaeghen:1999xj,Goloskokov:2005sd}. 
They describe well the high--energy data ($W > 10 \, \textrm{GeV}$,
HERA), where vector meson production couples mainly to the gluon 
GPD (see Fig.~\ref{fig:tslope}c) \cite{Frankfurt:1995jw}. 
The model of Ref.~\cite{Goloskokov:2005sd} 
also reproduces the $\phi$ production cross section data 
at lower energies, including CLAS \cite{Santoro:2008ai}
(see Fig.~\ref{fig:kroll}a). Interestingly,
none of the present calculations including quark GPDs can account for 
the fast rise of the $\rho^0$ production cross section below 
$W < 4 \, \textrm{GeV}$, seen in the recent CLAS \cite{Morrow:2008ek} 
and other data (see Fig.~\ref{fig:kroll}b). Whether this is 
due to shortcomings of the present quark GPD forms or inadequate 
modeling of higher--twist/finite--size corrections is presently 
under investigation. Model--independent tests of short--distance 
dominance through measurement of $t- (\bm{\Delta}_\perp^2-)$ slopes
(see above), as well as comparison of vector meson channels involving 
gluon and quark GPDs in different proportion ($\phi$: mostly gluons;
$\rho^0$: gluons and singlet quarks; $\rho^+$: non-singlet quarks,
no gluons) will help to distinguish between the two possibilities.
%
%

In the regime where small--size configurations dominate, meson 
production data can be used to extract information about the GPDs. 
The analysis of absolute cross sections is generally challenging, 
as the theoretical predictions depend on the detailed treatment 
of the hard scattering process (effective scale, higher--order QCD
corrections; see Ref.~\cite{Diehl:2007hd} for a recent discussion) 
as well as on the GPD parametrization/model.
One way of simplifying the analysis is to consider ratios of
cross sections in which the uncertainties associated with the
hard scattering process cancel, and one is left only with
ratios of the the GPDs or their integrals. An example is the
transverse target asymmetry in $\gamma^\ast_L p \rightarrow \rho^0 p$, 
which is sensitive to the Pauli form--factor type GPD $E$ figuring 
in the angular momentum sum rule \cite{Goeke:2001tz}. A similar
asymmetry can be studied in $\gamma^\ast_L p \rightarrow 
K^{\ast +} \Lambda$ by measuring the recoil polarization of the
produced $\Lambda$ \cite{Strikman:2008pi}; 
this process probes the $p \rightarrow \Lambda$
transition GPDs, which can be related to the usual flavor--diagonal
GPDs in the proton using $SU(3)$ flavor symmetry (see
Ref.~\cite{Strikman:2008pi} for other interesting ``ratio observables'').

Pseudoscalar meson production ($\pi, \eta, K$) in the small--size regime
probes the polarized GPDs in the nucleon. A particular feature of
$\pi^+$ (and to some extent also $K^+$) production is the existence
of a ``pole term'' in the GPD, in which the QCD operator measuring the
quark density is connected to the nucleon by $t$--channel $\pi^+ (K^+)$
exchange; it gives a contribution to the amplitude proportional to
the pion form factor, which is in fact the basis for measuring the
latter in electroproduction experiments \cite{Horn:2006tm}. 
Better insight into the relation between the ``pole'' and the 
``non--pole'' component of the GPD and their relative importance 
is necessary not only for improving the extraction of the pion form factor, 
but also for isolating the non-pole component related to the nucleon 
helicity structure. It could come \textit{e.g.}\ 
from model--independent comparisons of $\pi^+$ (which has a
pole) and $\pi^0$ (no pole) electroproduction data (see 
Ref.~\cite{Strikman:2008pi} for strange channels).
Experimental studies of exclusive pseudoscalar meson production are 
challenging because the $L/T$ virtual photon cross sections have to 
be separated by comparing data taken at different beam energies 
(Rosenbluth method). There are intriguing suggestions that $\sigma_T$
in exclusive pion production above the resonance region could be 
described as the limit of semi-inclusive production via the fragmentation 
mechanism \cite{Kaskulov:2008xc}; if confirmed, this could greatly aid 
the analysis of such processes.

\section{$\textrm{GPDs}$ in small--$x$ physics and $pp$ scattering}
\begin{figure}
\includegraphics[width=.7\textwidth]{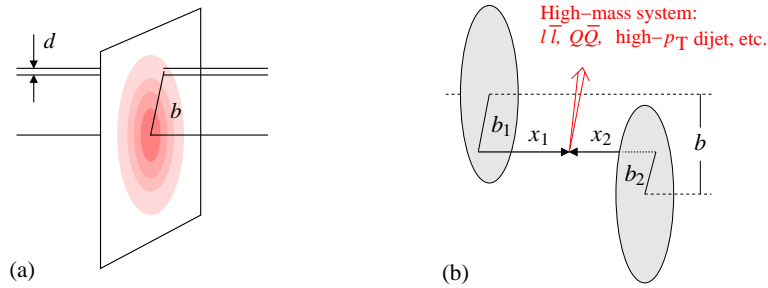}
\caption{(a) The dipole model of high--energy scattering.
The dipole--nucleon scattering amplitude at fixed impact parameter, $b$,
is proportional to the local gluon density in the transverse plane.
(b) Hard process in $pp$ scattering.} 
\label{fig:pp}
\end{figure}

The notion of the transverse spatial distribution of partons 
conveyed by the GPDs has many important applications in small--$x$ physics 
and high--energy $pp$ collisions with hard processes. Measurements
of the $t$--dependence of exclusive $J/\psi$ photo/electroproduction 
at HERA (\textit{cf.}\ Fig.~\ref{fig:tslope}b) and FNAL have provided 
a rather detailed picture of the transverse spatial distribution
of gluons with $10^{-4} < x < 10^{-2}$ \cite{Frankfurt:2005mc}. 
In particular, these experiments
have shown that the nucleon's gluonic transverse size at $Q^2 \sim 
\textrm{few GeV}^2$ is substantially smaller than its size in soft
hadronic interactions at high energies, and increases less rapidly
with energy (effective Regge slope $\alpha'_g \ll \alpha'_{\rm soft}$). 
This information provides essential input for theoretical studies of 
high--energy scattering in the QCD dipole model (see Fig.~\ref{fig:pp}a), 
where the dipole--nucleon interaction is governed by the local gluon 
density in the transverse plane,
and for modeling the initial condition of
non-linear QCD evolution equations describing the approach to the 
black--disk regime (unitarity limit).

In high--energy $pp$ scattering with hard processes (see Fig.~\ref{fig:pp}b), 
knowledge of the transverse spatial distribution of the partons in the
colliding protons allows one to calculate the probability 
of the hard process as a function of the $pp$ impact parameter, $b$,
and thus describe in detail the spectator interactions in such processes,
which are different from minimum bias events
\cite{Frankfurt:2003td}. This is particularly
important in hard diffractive $pp$ scattering, where the spectator
interactions determine the rapidity gap survival probability.
By measuring the $p_T$ dependence of central exclusive diffraction
$pp \rightarrow p + H + p \; (H = \textrm{high--mass system})$ one
can even extract information about the transverse spatial distribution
of gluons from the observed ``diffraction pattern'' \cite{Frankfurt:2006jp}.
%
%

\section{From densities to correlations}
A fast--moving hadron in QCD represents a complex many--body system, 
characterized by an equilibrium of creation and annihilation processes, 
capable of a wave function description as outlined by Gribov 
in the context of scalar field theory \cite{Gribov:1973jg}. 
The GPDs summarize the single--particle structure of this many--body 
system, at a transverse resolution scale where one can unambiguously 
identify quasi--free partons (\textit{i.e.}, integrated over transverse 
momenta up to a scale $\mu^2 \gg R_{\rm had}^{-2}$). However, there is
much more to be learned about this system than the 
single--particle densities. One example are the quantum
fluctuations of the parton densities, which result from the 
quantum--mechanical superposition of configurations of different size,
numbers of particles, \textit{etc.}\ in the wave function, and can
be revealed in hard diffractive $ep$ scattering \cite{Frankfurt:2008vi}. 
Another interesting
property are multi--parton correlations, which
are probed \textit{e.g.}\ in $pp$ scattering with multiple hard
processes. There are indications of significant transverse correlations
between partons, possibly related to the non--perturbative 
short--distance scale introduced by the spontaneous breaking of chiral 
symmetry in QCD (instanton size, ``constituent quark'' size) 
\cite{Frankfurt:2004kn}. These correlations show up as higher--twist 
corrections in inclusive DIS \cite{Sidorov:2006vu} and likely play 
an essential role in the $p_T$/azimuthal dependence
of semi--inclusive hadron production. Studying these correlations
should be the next step in the exploration of the partonic structure
of the nucleon.

Notice: Authored by Jefferson Science Associates, LLC under U.S.\ DOE
Contract No.~DE-AC05-06OR23177. The U.S.\ Government retains a
non-exclusive, paid-up, irrevocable, world-wide license to publish or
reproduce this manuscript for U.S.\ Government purposes.


\begin{thebibliography}{99}
%
%
\bibitem{Goeke:2001tz}
  K.~Goeke, M.~V.~Polyakov and M.~Vanderhaeghen,
  Prog.\ Part.\ Nucl.\ Phys.\  {\bf 47}, 401 (2001).
%
%
M.~Diehl,
Phys.\ Rept.\  {\bf 388}, 41 (2003).
%
%
  A.~V.~Belitsky and A.~V.~Radyushkin,
  Phys.\ Rept.\  {\bf 418}, 1 (2005).
%
%
\bibitem{Burkardt:2000za}
  M.~Burkardt,
  Phys.\ Rev.\  D {\bf 62}, 071503 (2000)
  [Erratum-ibid.\  D {\bf 66}, 119903 (2002)].
%
%
  P.~V.~Pobylitsa,
  Phys.\ Rev.\ D {\bf 66}, 094002 (2002).
%
%
  M.~Diehl,
  Eur.\ Phys.\ J.\ C {\bf 25}, 223 (2002).
%
%
\bibitem{Strikman:2003gz}
  M.~Strikman and C.~Weiss,
  Phys.\ Rev.\ D {\bf 69}, 054012 (2004);
%
%
  arXiv:0811.3631 [hep-ph].
%
%
\bibitem{Burkardt:2002hr}
  M.~Burkardt,
  Int.\ J.\ Mod.\ Phys.\  A {\bf 18}, 173 (2003).
%
%
\bibitem{Diehl:2005jf}
  M.~Diehl and Ph.~Hagler,
  Eur.\ Phys.\ J.\  C {\bf 44}, 87 (2005).
%
%
\bibitem{Ji:1996ek}
  X.~D.~Ji,
  Phys.\ Rev.\ Lett.\  {\bf 78}, 610 (1997);
%
%
  Phys.\ Rev.\ D {\bf 55}, 7114 (1997).
%
%
\bibitem{Voutier:2009sc}
  For a summary, see:
  E.~Voutier,
  these proceedings,
  arXiv:0901.3016 [nucl-ex].
%
%
\bibitem{Polyakov:2002yz}
  M.~V.~Polyakov,
  Phys.\ Lett.\  B {\bf 555}, 57 (2003).
%
%
\bibitem{Goeke:2007fp}
  K.~Goeke 
  {\it et al}, 
  Phys.\ Rev.\  D {\bf 75}, 094021 (2007).
%
%
\bibitem{Hagler:2007xi}
  Ph.~Hagler {\it et al.}  [LHPC Collaborations],
  Phys.\ Rev.\  D {\bf 77}, 094502 (2008).
%
%
\bibitem{Zanotti:2008zm}
  For a recent review, see: J.~M.~Zanotti,
  arXiv:0812.3845 [hep-lat].
%
%
\bibitem{Munoz Camacho:2006hx}
  C.~Munoz Camacho {\it et al.} 
  [JLAB Hall A \& Hall A DVCS Coll.],
  Phys.\ Rev.\ Lett.\  {\bf 97}, 262002 (2006).
%
%
\bibitem{Polyakov:2002wz}
  M.~V.~Polyakov and A.~G.~Shuvaev,
  arXiv:hep-ph/0207153.
%
%
\bibitem{SemenovTianShansky:2008mp}
  K.~M.~Semenov-Tian-Shansky,
  Eur.\ Phys.\ J.\  A {\bf 36}, 303 (2008).
%
%
  M.~V.~Polyakov and K.~M.~Semenov-Tian-Shansky,
  arXiv:0811.2901 [hep-ph].
%
%
\bibitem{Mueller:2005ed}
  D.~Mueller and A.~Schafer,
  Nucl.\ Phys.\  B {\bf 739}, 1 (2006).
%
%
\bibitem{Radyushkin:1997ki}
  A.~V.~Radyushkin,
  Phys.\ Rev.\  D {\bf 56}, 5524 (1997);
%
%
  D {\bf 59}, 014030 (1999).
%
%
\bibitem{Kumericki:2007sa}
  K.~Kumericki, D.~Mueller and K.~Passek-Kumericki,
  Nucl.\ Phys.\  B {\bf 794}, 244 (2008).
%
%
\bibitem{Guzey:2008ys}
  V.~Guzey and T.~Teckentrup,
  Phys.\ Rev.\  D {\bf 79}, 017501 (2009).
%
%
\bibitem{Polyakov:2008xm}
  M.~V.~Polyakov and M.~Vanderhaeghen,
  arXiv:0803.1271 [hep-ph].
%
%
\bibitem{Teryaev:2005uj}
  O.~V.~Teryaev,
  arXiv:hep-ph/0510031.
%
%
  I.~V.~Anikin and O.~V.~Teryaev,
  Phys.\ Rev.\  D {\bf 76}, 056007 (2007).
%
%
  M.~Diehl and D.~Y.~Ivanov,
  Eur.\ Phys.\ J.\  C {\bf 52}, 919 (2007).
%
%
\bibitem{Polyakov:1999gs}
  M.~V.~Polyakov and C.~Weiss,
  Phys.\ Rev.\  D {\bf 60}, 114017 (1999).
%
%
\bibitem{Frankfurt:2005mc}
  L.~Frankfurt, M.~Strikman and C.~Weiss,
  Ann.\ Rev.\ Nucl.\ Part.\ Sci.\  {\bf 55}, 403 (2005).
%
%
\bibitem{Frankfurt:1995jw}
  L.~Frankfurt, W.~Koepf and M.~Strikman,
  Phys.\ Rev.\  D {\bf 54}, 3194 (1996).
%
%
\bibitem{Vanderhaeghen:1999xj}
  M.~Vanderhaeghen, P.~A.~M.~Guichon and M.~Guidal,
  Phys.\ Rev.\  D {\bf 60}, 094017 (1999).
%
%
\bibitem{Goloskokov:2005sd}
  S.~V.~Goloskokov and P.~Kroll,
  Eur.\ Phys.\ J.\  C {\bf 42}, 281 (2005);
%
%
  C {\bf 50}, 829 (2007);
%
%
  C {\bf 53}, 367 (2008).
%
%
\bibitem{Santoro:2008ai}
  J.~P.~Santoro {\it et al.}  [CLAS Collaboration],
  Phys.\ Rev.\  C {\bf 78}, 025210 (2008).
%
%
\bibitem{Morrow:2008ek}
  S.~A.~Morrow {\it et al.}  [CLAS Collaboration],
  arXiv:0807.3834 [hep-ex].
%
%
\bibitem{Diehl:2007hd}
  M.~Diehl and W.~Kugler,
  Eur.\ Phys.\ J.\  C {\bf 52}, 933 (2007).
%
%
\bibitem{Strikman:2008pi}
  M.~Strikman and C.~Weiss,
  arXiv:0804.0456 [hep-ph].
%
%
\bibitem{Horn:2006tm}
  T.~Horn {\it et al.}
  [Jefferson Lab F(pi)-2 Collaboration],
  Phys.\ Rev.\ Lett.\  {\bf 97}, 192001 (2006).
%
%
  T.~Horn {\it et al.},
  arXiv:0707.1794 [nucl-ex].
%
%
  H.~P.~Blok {\it et al.},
  Phys.\ Rev.\  C {\bf 78}, 045202 (2008).
%
%
  G.~M.~Huber {\it et al.},
  Phys.\ Rev.\  C {\bf 78}, 045203 (2008).
%
%
\bibitem{Kaskulov:2008xc}
  M.~M.~Kaskulov, K.~Gallmeister and U.~Mosel,
  Phys.\ Rev.\  D {\bf 78}, 114022 (2008).
%
%
\bibitem{Frankfurt:2003td}
  L.~Frankfurt, M.~Strikman and C.~Weiss,
  Phys.\ Rev.\  D {\bf 69}, 114010 (2004).
%
%
\bibitem{Frankfurt:2006jp}
  L.~Frankfurt, C.~E.~Hyde, M.~Strikman and C.~Weiss,
  Phys.\ Rev.\  D {\bf 75}, 054009 (2007).
%
%
\bibitem{Gribov:1973jg}
  V.~N.~Gribov,
  arXiv:hep-ph/0006158.
%
%
\bibitem{Frankfurt:2008vi}
  L.~Frankfurt, M.~Strikman, D.~Treleani and C.~Weiss,
  Phys.\ Rev.\ Lett.\  {\bf 101}, 202003 (2008).
%
%
\bibitem{Frankfurt:2004kn}
  L.~Frankfurt, M.~Strikman and C.~Weiss,
  Annalen Phys.\  {\bf 13}, 665 (2004).
%
%
\bibitem{Sidorov:2006vu}
  A.~V.~Sidorov and C.~Weiss,
  Phys.\ Rev.\  D {\bf 73}, 074016 (2006).
%
%
\end{thebibliography}
\end{document}